\documentclass[conference,compsoc]{IEEEtran}
\ifCLASSOPTIONcompsoc
  \usepackage[nocompress]{cite}
\else
  \usepackage{cite}
\fi
\ifCLASSINFOpdf
\else
\fi
\usepackage{dirtytalk}
\usepackage{hyperref}
\usepackage{graphicx}

\IEEEoverridecommandlockouts 
\begin{document}

\date{}

\title{Analyzing Trends in Tor}

\author{\IEEEauthorblockN{Chaitanya Rahalkar}
\thanks{All authors have contributed equally}
\IEEEauthorblockA{School of Computer Science\\
Georgia Institute of Technology\\
cr@gatech.edu}
\and
\IEEEauthorblockN{Anushka Virgaonkar}
\IEEEauthorblockA{School of Computer Science\\
Georgia Institute of Technology\\
avirgaonkar3@gatech.edu}
\and
\IEEEauthorblockN{Kethaki Varadan}
\IEEEauthorblockA{School of Computer Science\\
Georgia Institute of Technology\\
kvaradan3@gatech.edu}
}

\maketitle

\begin{abstract}
The Tor Network has been a significant part of the Internet for years. Tor was originally started in the Naval Research Laboratory for anonymous Internet browsing and Internet-based communication. From being used for anonymous communications, it has now segmented into various other use-cases like censorship circumvention, performing illegal activities, etc. In this paper, we perform empirical measurements on the Tor network to analyze the trends in Tor over the years. We gather our measurements data through our measurement scripts, past research in this domain, and aggregated data provided by the Tor metrics directory. We use this data to analyze trends and understand the incidents that caused fluctuations in the trends of different data parameters. We collect measurements data for Tor parameters like Tor users, onion services, Tor relays, and bridges, etc. We also study censorship-related events and study trends by analyzing censorship-related metrics. Finally, we touch upon the location diversity in Tor and study how the Tor circuit selection and construction are impacted by the bandwidth distribution of Tor relays across geographies.
\end{abstract}

\section{Introduction}

Tor is a circuit-based low-latency anonymous communication service \cite{dingledine04} \cite{isaacson13}. Over the years, the usage of Tor has increased and it is being used in a variety of scenarios, both good and bad (legal and illegal as well). The current trends show that Tor is suitable due to its relatively low latency for being used in circumventing censorship and as a medium for hiding online illegal activity \cite{jardine20}. However, there are many cases where users strictly use Tor for protecting their private data as it is transmitted over the internet. It is used as a privacy-enhancing technique by privacy advocates, journalists, activists, law enforcement and is also for censorship circumvention purposes. Studying the usage patterns in Tor is critical to get a realistic and precise overview of the Tor infrastructure in terms of its privacy and anonymity guarantees. It leads us to the path for understanding the future updates and changes that Tor would undergo. 
\\
Tor achieves its anonymity guarantees by sending traffic in circuits. A circuit consists of an entry node, a middle node, and an exit node. The entry node knows the sender and encrypts the data packet before sending it to the middle node. The middle node cannot decrypt the packet to reveal the identity of the sender. The middle node encrypts the data packet further and then sends it to the exit node. The exit node decrypts the received packet to identify the receiver. This decryption does not reveal the identity of the sender. Tor's anonymity guarantee dictates that at any given node, both the sender and receiver of a packet cannot be identified. 
\\
Relays in the Tor network are run by volunteers. They provide the Tor directory servers with information about their relays such as available bandwidth, IP address, exit policies. The directory authorities maintain records of the relays by storing their information and tracking their statuses. A consensus is reached for agreeing upon the list of relays that would be utilized in the Tor network. A user who wishes to use the Tor network installs the Tor client on their device. The Tor client fetches the information about the relays by downloading the consensus file from the directory authorities or their mirrors. The consensus file is refreshed hourly to reflect updated network information. To form a circuit, the Tor client selects three relays. The exit relay is chosen first based on their exit policies. An exit policy specifies the ports that the relay provides for communication. The entry relay is chosen next by the Tor circuit. It is selected based on reliability and bandwidth. Finally, the middle node is chosen. The relay selection in a Tor circuit is a random process with a bias towards higher bandwidth relays to ensure high performance. \\
Once the client picks the relays for the path, it builds a circuit, which consists of three cryptographic tunnels, one between the client and each of the relays. The client first contacts the entry and uses an authenticated Diffie-Hellman handshake to share a secret key with it. The client and the entry use this secret key to encrypt and authenticate all of their communications, creating a secure tunnel between them. Through this tunnel, the client then asks the entry to extend the circuit to the middle. The client uses the same protocol to establish a secure tunnel with the middle that is layered inside of the tunnel between the client and the entry. Finally, the client uses the same protocol to extend the circuit to the exit through the tunnels to the entry and the middle. When the circuit is completed, then the client can attach a stream to it by opening a SOCKS connection. The stream’s data is sent over the circuit and encrypted multiple times in a telescopic fashion such that no one on the path can link the source of the data to its destination \cite{wacek13}\cite{tormanual}.
\\
Studying anonymous networks poses various ethical and legal challenges. However, finding various parameters in an anonymous network like Tor is critical to understanding its security posture and threat landscape. We follow up on the study conducted in McCoy \textit{et al.} which studied how Tor was being used and misused, and who used Tor \cite{mccoy08}. We present current results of Tor usage by first studying the evolution of Tor since its inception. We use the data collected and managed by Tor metrics to quantify the evolutionary changes. 
\\
We take several ethical measures in conducting our study. We do not host a router and collect data that would contain the user's private data at any point in time. We rely on existing data sources and research. The major source that we rely on is the official Tor metrics directory which uses data that is collected through statistical aggregation for preserving privacy. The tools that we run to collect present data do not involve packet inspection which results in revealing private data. 

\subsection{Evolution of Tor}
The Tor network is based on onion routing, the communication technique developed in the 1990s which aimed to ensure both private and anonymous communication. Messages are first encrypted several times before being sent across nodes (onion routers) in the network that serves to successively decrypt and pass along the encrypted message object. Anonymity is guaranteed in this process since every node knows only the previous and successive nodes and the state of the received message at any time, while encryption ensures that privacy is preserved (although the system is still vulnerable to other types of attacks). Tor was developed in the early 2000s and improved upon early limitations identified with onion routing, and the first tools to implement it \cite{dingledine04}. 

In 2008, the Tor browser was deployed to enable easier and more widespread use of the Tor network. Tor has since been noted for its impact on societal and sociopolitical causes: as a key communication tool during the Arab Spring uprisings, for instance, or to actively evade censors like China’s Great Firewall. Improvements in Tor have been made in anticipation and as a consequence of its use in these specific scenarios \cite{tschantz16}. Thus, Tor provides an interesting landscape to examine developments over time: from its origins in academic cryptographic research to its current position as the most widely known and usable tool to protect privacy and anonymity as much as possible.
\section{Related Work}\label{relwork}
Previous measurement studies have examined Tor in some depth. In one of the first such in 2008, McCoy \textit{et al.} participated in the Tor network by setting up a relay node and analyzing through traffic \cite{mccoy08}. Broadly, the authors aimed to understand how Tor was being used \textit{and} misused, and who was using it. The authors characterized the different types of traffic, distinguished between secure and insecure protocols being used, and examined the geopolitical distributions of clients and routers. 

While in-depth, this study raised questions regarding the ethics of data collection and privacy \cite{mccoy08}. For example, the author logged enough data, including IP addresses, to check application-level headers of traffic exiting their router. More recent studies, whereas, have specifically used tools and techniques based on privacy-preserving advancements developed since then \cite{jansen16}\cite{wails19}. While we examined Tor’s usage and evolution following the same questions raised by McCoy \textit{et al.}, our measurement approach differs considerably.

\section{Approach}
\begin{figure}[hbt!]
\centering
\includegraphics[scale=0.4]{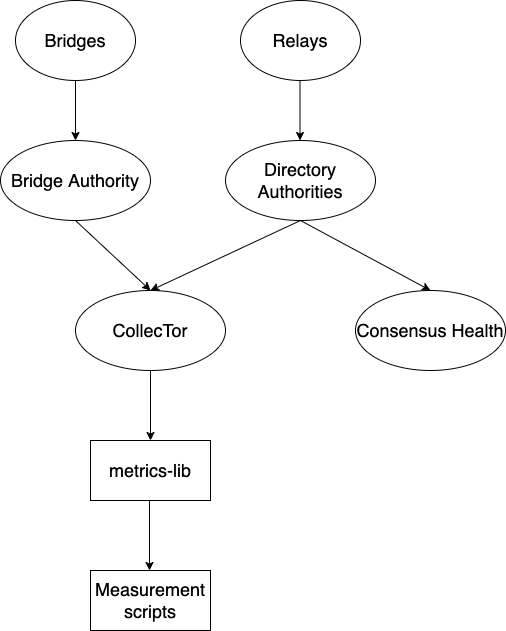}
\caption{Gathering Measurement Data using Tor Libraries}
\label{fig:approach}
\end{figure}

The main challenge we faced was gathering measurement data without actually participating in the network by hosting a relay or bridge. Instead, we relied on data collected from three different sources:
\begin{enumerate}
    \item Measurements obtained directly 
    \item Measurements taken from existing research 
    \item Measurements provided by the Tor metrics directory
\end{enumerate}
The Tor Project specifically gathers metrics information using a Bridge Authority, a special-purpose relay that maintains a list of bridges. Directory authorities are other special-purpose relays that maintain a list of all currently-running relays, and periodically publish the consensus. All metrics information from these sources is aggregated and collected at CollecTor, a centralized directory\cite{collecting}. Tor's metrics-lib is a Java API for interacting with and parsing data from CollecTor. Stem, another Tor controller library, enables interacting with the Tor network using scripts.\cite{stem}. Our direct measurements are Python scripts written against the Tor metrics and Stem libraries (see \autoref{fig:approach}).

Where direct measurement seemed impractical, we relied on data obtained from existing research. Since we aimed to examine trends in the Tor network over the years, we needed to analyze historical data. For this, we used data gathered over the past fifteen years that is provided by the Tor metrics directory. All historical trends presented in this paper are produced from this data. We used Matplotlib, a Python library, for data visualization and analysis \cite{hunter2007matplotlib}.

\section{Trends in Tor Users}\label{usertrends}

\begin{figure}[hbt!]
\centering
\includegraphics[scale=0.4]{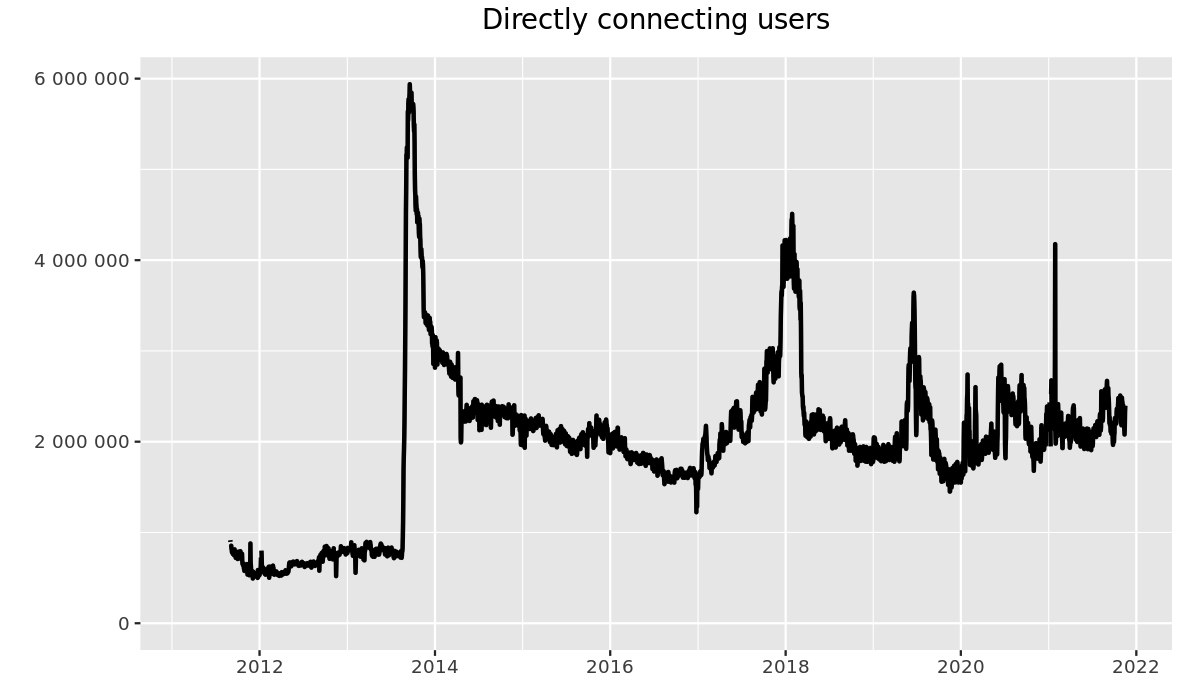}
\caption{Number of users directly connecting to Tor over the period from 01/01/2000 to 11/18/2021.}
\label{fig:usersfig}
\end{figure}

\autoref{fig:usersfig} displays the number of users connecting to Tor over the past twenty years. The general increase in the number is expected but is punctuated by sharp spikes at certain periods. We examine the first such significant increase observed around August 2013, where the number of users doubled to 2 million and then 4 million before peaking to ~6 million in September. Many news outlets considered the uptick a consequence of the Snowden revelations of July 2013 which exposed mass surveillance by the NSA\cite{dredge13}\cite{isaacson13}. This underscores Tor’s use and perception as a tool for protecting privacy. Security firm, Fox IT, however, more plausibly attributed the increase to a massive click-fraud botnet Mevade\cite{foxit13}. 

Mevade operated using a Tor hidden service for its command and control (C\&C) operations. Hidden services only accessible to Tor, were introduced to provide complete anonymity in communication. The public encryption key of the hidden servers is published on Tor's directory servers. An onion pseudo-domain is generated from the keys for the service which is used to resolve the hidden service and to establish communication with it. The actual IP address of the servers remains hidden and identifying the server from the \texttt{.onion} domain is not possible. Since Mevade used a hidden service effectively concealing its location, it was resistant to censor takedown efforts, and since the nature of the C\&C was concealed, the detection was also difficult. Tor released an update to include a new handshake that was prioritized over the older handshake by Tor relays. This mitigation relied on the assumption that the Mevade botnet would not update their Tor component and therefore, their connections would not be reestablished. In a public statement, the Tor developers acknowledged this solution for not being very resilient in stopping botnet attacks on Tor and called for the help of researchers for designing a solution to the problem. The use of Tor's hidden services for concealed botnet C\&C first started with Skynet in 2012. Skynet operated to conduct DDoS attacks, Bitcoin mining, and banking. Previous studies\cite{nicholashopper} have studied various strategies for making the use of Tor for botnet C\&C infeasible and unreliable for botnet operators. However, some strategies demand protocol changes in Tor and prove to be impractical to deploy. The use of Tor in botnets is still prevalent and the need for practical solutions persists.

\begin{table*}[]
	\begin{center}
		\begin{tabular}{p{0.2\textwidth}|p{0.2\textwidth}}
			\textbf{Country} &
			\textbf{Mean Daily Users} \\ \hline
			\textbf{United States} &  471213 \\ 
			\textbf{Russia} &  323847 \\ 
			\textbf{Germany} &  180257 \\ 
			\textbf{Netherlands} &  104542 \\ 
			\textbf{France} &  71036 \\ 
			\textbf{Indonesia} &  64245 \\ 
			\textbf{United Kingdom} &  61182 \\ 
			\textbf{Ukraine} &  59369 \\ 
			\textbf{India} &  52048 \\ 
			\textbf{Lithuania} & 45873
		\end{tabular}
	\end{center}
	\caption{Top 10 countries by mean number of daily users over the period - 08/25/2020 to 11/24/2021}
	\label{userstbl}
\end{table*}

\autoref{userstbl} shows the top 10 countries by mean daily users using Tor. It is interesting to note that though there are no relays present in Russia, it observes the second-most mean daily users count. This depicts the inequality in the source of Tor traffic and the location of infrastructure required for Tor.

\section{Trends in Onion Services}

\begin{figure}[hbt!]
\centering
\includegraphics[scale=0.4]{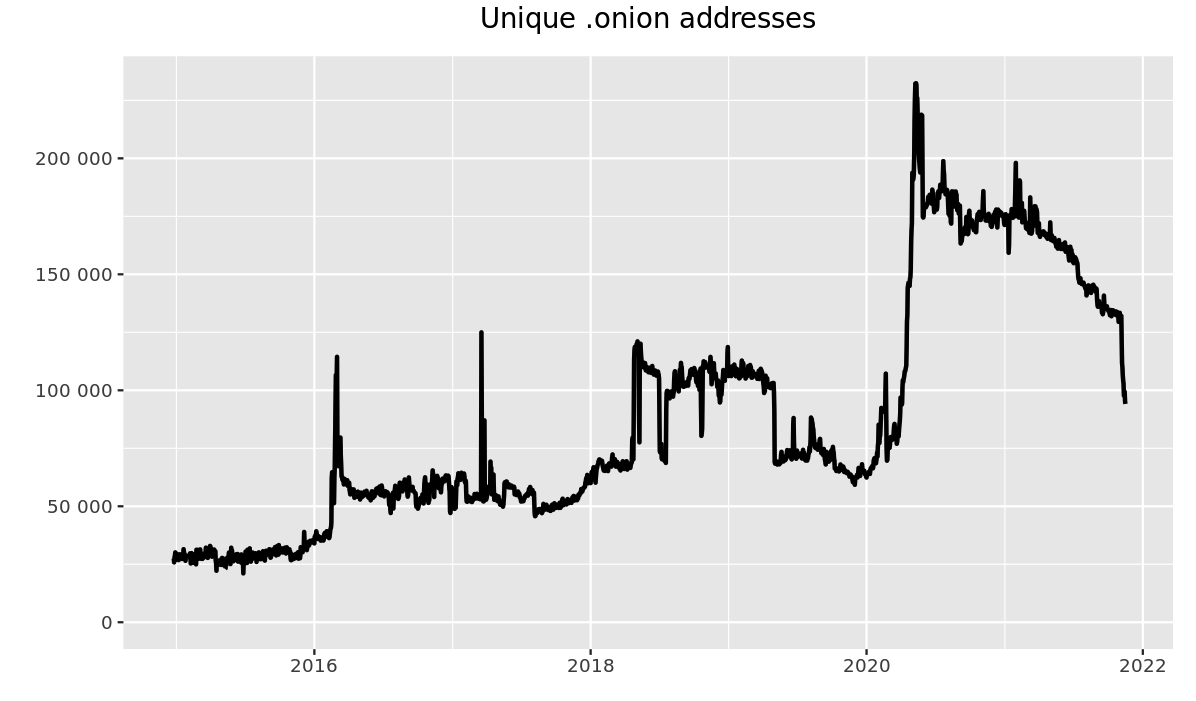}
\caption{Onion Server Statistics}
\label{fig:torstats}
\end{figure}

Onion services are anonymous services accessible only through the Tor network. Because all traffic to the services is encrypted end-to-end, the use of onion services provides privacy and anonymity guarantees. However, they remain susceptible to other deanonymization attacks. Domain ownership is difficult to identify because the onion addresses of services are dynamically generated and do not require purchasing a domain name. As seen in \autoref{usertrends}, botnets have exploited this for concealment. To similar ends, censorship circumvention tools also make use of these services.

From \autoref{fig:torstats} there is a general increase in the number of onion addresses over the years, but with sharp spikes at some periods. The most recent spike was in early to mid-2020, where the number of sites tripled from around 75,000 to 300,000 in the space of a few months. One explanation for the increase could be Tor's updates around October 2020, where support for v2 addresses (16 character domains) was removed in favor of v3 addresses, the more-secure 56-character domains. This could be explained by the creation of a large number of the new v3 addresses, while the previous v2 addresses were still present since support for the latter wasn't removed until 2021.   

\section{Trends in Tor Application Information}
In this section, we analyze the trends in the Tor browser application downloads and update requests. This metric indicates how many times the Tor browser was downloaded from the Tor website (all means of downloading the browser - direct download, mirror links, cloud storage links, etc.). This metric only considers only full download attempts of the browser i.e. attempts where the browser was fully downloaded by the client. These download metrics are categorized by platforms (MacOS, Linux, Windows) for which the browser was downloaded. "Update Pings" are requests made by the Tor browser to check whether a new version of the browser is available or not. By studying the Tor Browser download information, we can understand how many of the Tor users are actual browser-based users. Since there are other ways to be a part of the Tor network (using a different browser with a Tor proxy, other Tor browsers, Tor scripts, Tor relays and bridges, etc.), this metric segregates regular Tor users (that depend on the Tor browser), from the other types. Looking at the download trends, we see that Tor has been largely downloaded by Windows users as compared to MacOS and Linux. 

\begin{figure}[hbt!]
\centering
\includegraphics[width=0.4\paperwidth]{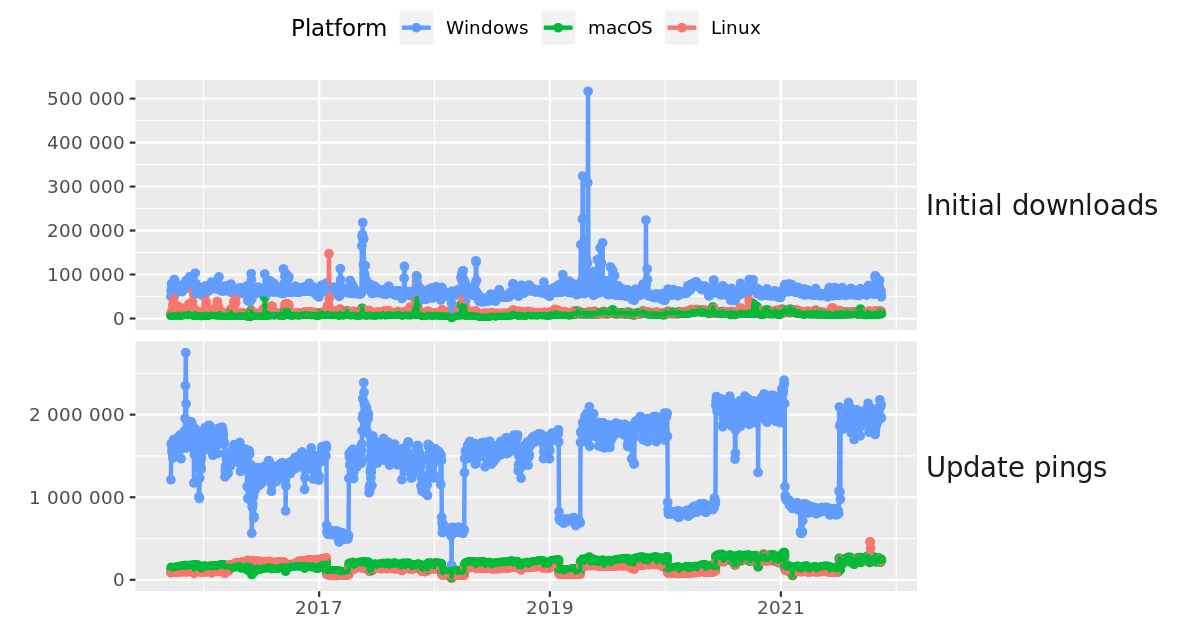}
\caption{Tor Application Information}
\label{fig:torappinfo}
\end{figure}

\section{Trends in Tor Relays and Bridges}

\begin{figure}[hbt!]
\centering
\includegraphics[scale=0.4]{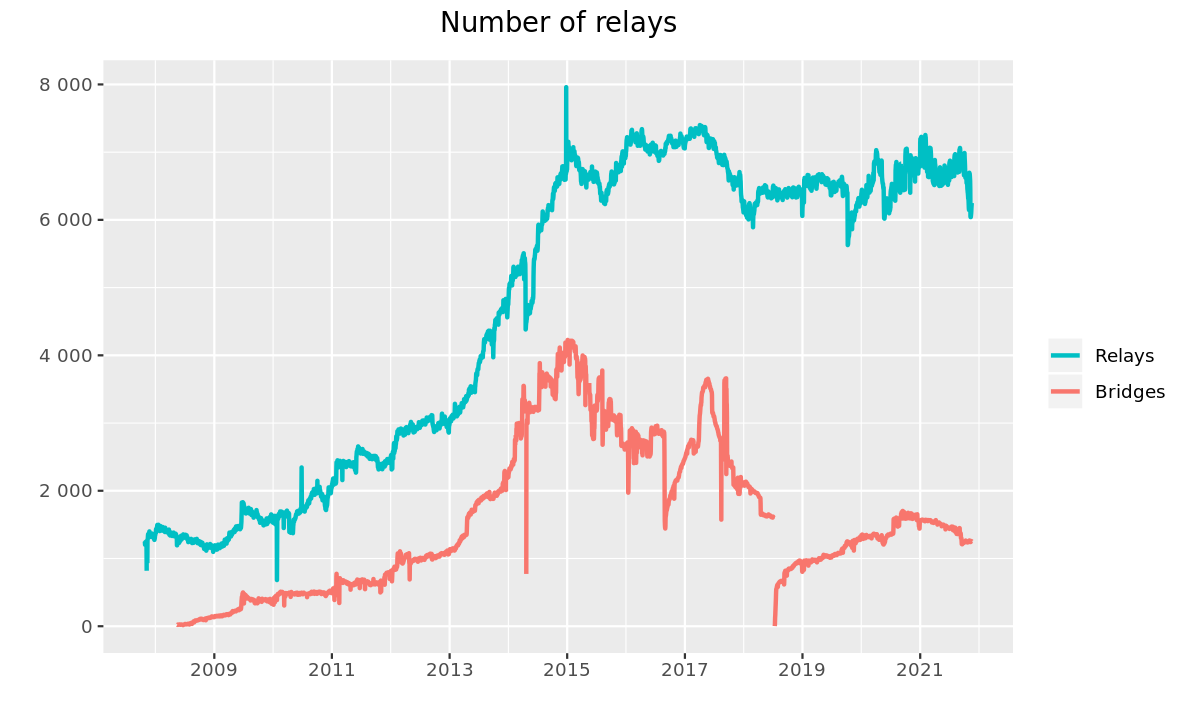}
\caption{Relays and Bridges Statistics}
\label{fig:relaybridgestats}
\end{figure}

\autoref{fig:relaybridgestats} shows the trend in Tor relays and bridges over the span of of 12 years. The bridge data is missing for roughly 3 months which is why the graph has a disconnected line. 

A Tor Relay is a publicly-listed node in the Tor network that forwards traffic on behalf of clients, and that registers itself with the directory authorities. Tor Bridges are essentially Tor relays that are not listed in the public Tor directory \cite{whatisbridge}. From the graph, we see that the number of Tor bridges has been decreasing over the years, in the Tor network. In 2021, Tor announced that it is giving away swag to anyone who runs a Tor bridge. The dire need for a large number of Tor bridges is so that most of them would be prevented from being blacklisted. The graph shows that the number of bridges and relays was at an all-time high during 2015. This can be accounted to a large number of high-bandwidth bridges powered by Amazon and Google being rate-limited. Consequently, the two bridges that were considered the backbone of the system were being replaced by corporations and users hosting their own bridges and relays. This was also the time (specifically February of 2015) when Meek (a highly reliable and secure Pluggable Transport) was officially supported in Orbot (Android-based Tor browser) and Tor, which opened doors for censorship circumvention in many countries.

\section{Trends in Tor Relay Bandwidth}
\begin{figure}[hbt!]
\centering
\includegraphics[width=0.4\paperwidth]{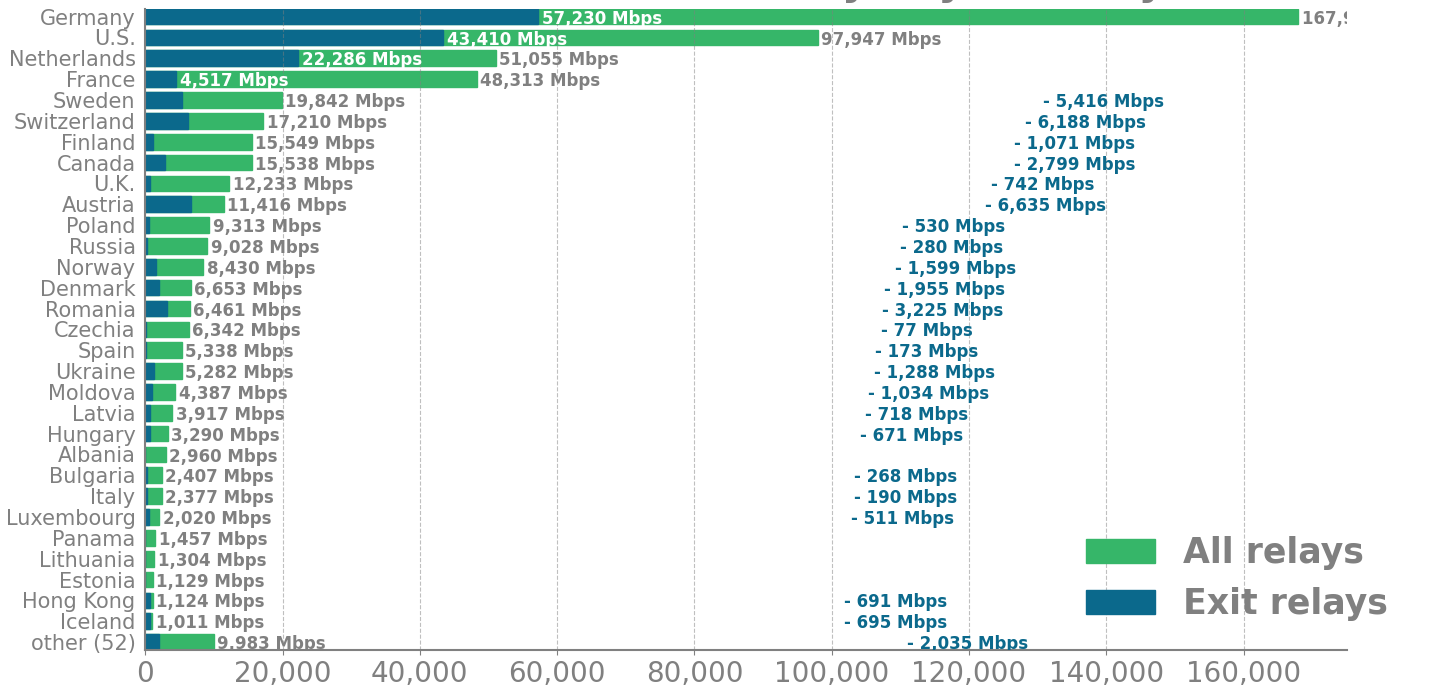}
\caption{Relay Bandwidth Statistics}
\label{fig:relaybw}
\end{figure}

In this section, we study the trends in the Tor relay bandwidth distribution across different countries. Studying the bandwidth distribution across geographies is important because it allows us to understand how the Tor circuit selection process would take place. This analysis would help us understand the location diversity in Tor which is explained in the later sections. \autoref{fig:relaybw} shows the bandwidth distribution of Tor relays and more specifically Tor Exit relays, by country. A Tor circuit comprises of three nodes - the entry, middle, and exit node. During the circuit selection process, the Tor network considers the network bandwidth capacity of the nodes to ensure good performance. By studying the network bandwidth capacities of the Tor relays by country, we can get a high-level understanding of which Tor relay would be typically picked during the circuit selection process. As shown in the figure, Germany has the highest relay bandwidth capacity. Therefore, in a typical circuit selection process, there's a high chance for Germany's node to be selected since we generically consider that a given node in Germany would have considerably more bandwidth as compared to nodes from other countries. We performed an experiment where the Tor circuit selection process was emulated using the Python STEM library. We ran the circuit emulation process for 10,000 iterations. In the circuit emulation process, we were able to get only the first five countries. This is because the circuit selection process considers the location of the client and several other factors. Since the experiments were running from the United States, getting a close-by node with good bandwidth capability was highly probable and this factor overwhelmed the possibility of getting a node from a lower-ranked country. The results of our observations were consistent with what we can reason from the figure. \autoref{tab:node-selection} shows our observations, ranked by country (top 5 in the graph are consistent with the five observed in the circuit selection process)- 

\begin{table}[]
\begin{tabular}{l|l|l|l}

Country & Entry Node & Middle Node & Exit Node \\
\hline
Germany & 5237 & 4884 & 5727 \\
United States & 3113 & 3733 & 3263 \\
Netherlands & 1199 & 1210 & 824 \\
France & 451 & 173 & 186
\end{tabular}
\caption{Tor Node Selection Ranked By Country}
\label{tab:node-selection}
\end{table}


\section{Censorship}
\begin{figure}[hbt!]
\centering
\includegraphics[scale=0.4]{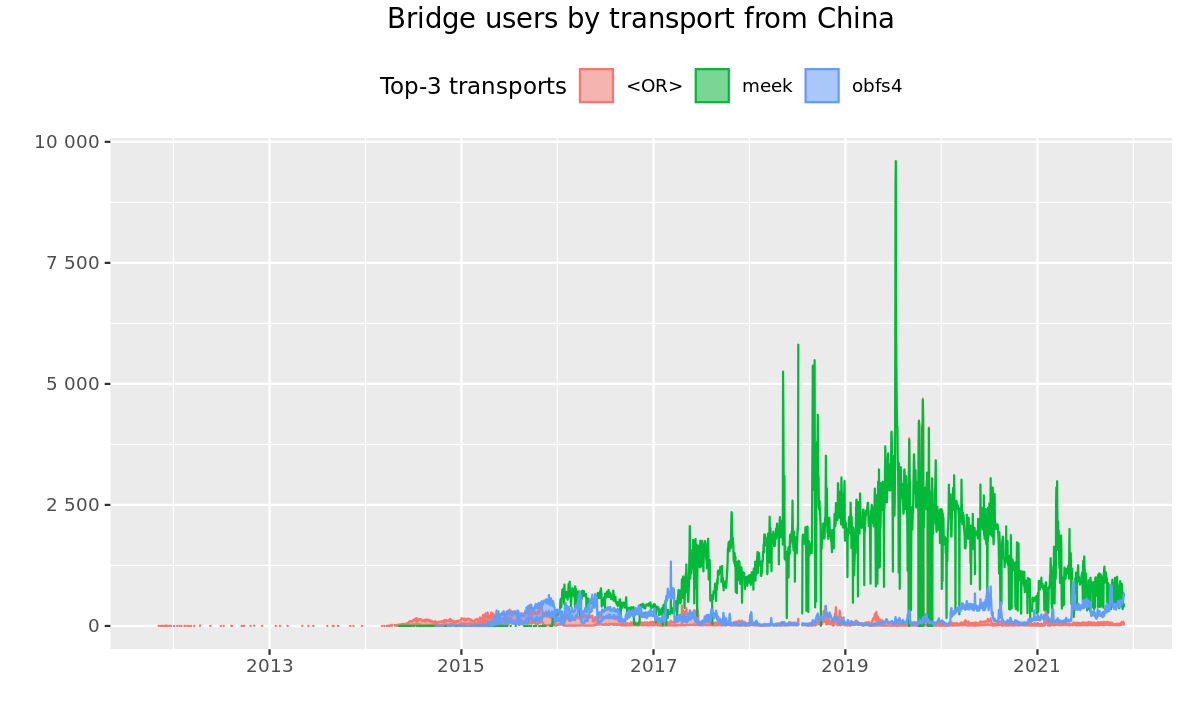}
\caption{Pluggable transports in China}
\label{fig:china}
\end{figure}

Tor is widely used as a medium to circumvent censorship. Access to certain kinds of websites that contain propaganda messages, pornographic content, social media sites, etc. is the kinds of sites that are typically censored if the content is not authoritatively approved in the country. Over the years, governments have been aware of Tor and the methodologies it uses to perform censorship circumvention. Therefore, such countries often make attempts to restrict users from accessing censored content via Tor. Countries like China have been known to restrict access to the Tor network and have kept up with the recent technological advancements that the Tor project is implementing to make Tor more accessible to users where Tor is censored. The Tor project introduced Pluggable transports in the Tor network to prevent such censorship attempts. However, China has managed to detect and block many of the pluggable transport techniques over the year. \autoref{fig:china} shows the trend in the number of bridge users that use different pluggable transports in China to access the Tor network. As seen in the graph, the $\langle OR \rangle$ and obfs4 pluggable transport mechanism to access the network has suppressed/reduced over the years (with sudden spikes at some points). This is because China has found ways to suppress these mechanisms. However, the meek pluggable transport mechanism is still widely used. This is because the meek pluggable transport obfuscates the Tor traffic in TLS which makes it much harder to detect. Also, the Meek-supported pluggable transport bridges are hosted by AWS and Azure in China. Blocking the Meek pluggable transport would require them to block the AWS bridges which would consequently result in the blockage of the entire AWS and Azure IP range. Many Chinese companies heavily rely on AWS and Azure to host their cloud-based infrastructure. 

\begin{center}
  
\end{center}

\section{Location diversity in Tor}

\begin{table*}[hbt!]
\begin{tabular}{l|l|l|l|l|l|l|l|l|l}

\textbf{Country} &
 \textbf{\begin{tabular}[c]{@{}l@{}}Autonomous\\ System\end{tabular}} &
  \textbf{\begin{tabular}[c]{@{}l@{}}Consensus\\ Weight\end{tabular}} &
  \textbf{\begin{tabular}[c]{@{}l@{}}Advertised\\ Bandwidth\end{tabular}} &
  \textbf{\begin{tabular}[c]{@{}l@{}}Guard\\ Probability\end{tabular}} &
  \textbf{\begin{tabular}[c]{@{}l@{}}Middle\\ Probability\end{tabular}} &
  \textbf{\begin{tabular}[c]{@{}l@{}}Exit\\ Probability\end{tabular}} &
  \textbf{Relays} &
 \textbf{Guard} &
  \textbf{Exit} \\ \hline
\textbf{Germany} &
  111 &
  34.8125\% &
  20463.55 MiB/s &
  33.8910\% &
  31.0940\% &
  39.6611\% &
  1459 &
  786 &
  319 \\
\textbf{United States} &
  185 &
  13.5932\% &
  11720.78 MiB/s &
  10.3794\% &
  10.6811\% &
  19.9947\% &
   1632 &
   948 &
   564 \\ 
\textbf{Netherlands} &
   78 &
   10.9938\% &
   6179.12 MiB/s &
   8.2127\% &
   8.5996\% &
       16.4018\% &
       402 &
       260 &
       129 \\ 
\textbf{France} &
      32 &
      9.7469\% &
      5877.89 MiB/s &
      13.3639\% &
      12.8505\% &
          2.7240\% &
          440 &
          281 &
          40

\end{tabular}
\caption{Relay probabilities in top 4 countries who have the highest probability of selection of relays in Tor circuits}
\label{relayprb}
\end{table*}

The Tor client selects relays to construct a Tor circuit. For selection, a relay with a higher consensus weight is given priority. The consensus weight is the bandwidth advertised by the relay, which is included in the hourly consensus published by the Tor directory authorities. 
\autoref{relayprb} shows that the relays present in Germany have the highest probability for being selected as guard, middle, and exit nodes, which is because the relays located in Germany have the highest advertised bandwidth which contributes towards their high consensus weight. Therefore, the probability of selecting relays for inclusion in a Tor circuit is the highest when the relays are present in Germany. That means, among all the available relays for circuit selection, the relays located in Germany hold the highest chance of selection in a Tor circuit. Furthermore, this probability distribution is highly concentrated in 4 countries: Germany, the United States, Netherlands, and France. 

We observe that even though the United States has the highest numbers of relays present, their guard, middle and exit probabilities are less than the relays located in Germany. This is because the advertised bandwidth of the former is less than that of the latter, i.e. that the bandwidth is disproportionately distributed in relays present in Germany. This was also observed by McCoy \textit{et al.} in 2008: while most Tor routers were located in both Germany and the United States, Germany alone contributed to nearly half of the network’s total bandwidth (45\% to the United States' 23\%) \cite{mccoy08}. We conclude that, even over a decade later, the bandwidth distribution has not spread out to include other geographical regions. This has an impact on Tor's location diversity. 
Anonymous networks like Tor are susceptible to eavesdropping attacks that involve linking a sender and receiver through packet counting or timing attacks.
This harms the privacy and anonymity goals of the network. In a Tor routing circuit, traffic flows from the source to the entry node, from the entry node to the middle node, from the middle node to the exit node, and from the exit node to the destination. If at any point in this circuit, traffic flows from any sender-receiver pair through the same autonomous system more than once, then that autonomous system can correlate the traffic it observes to infer the sender-receiver pair. Previous studies focused on onion routing have proven that an adversary observing c of the n nodes in the network can break $(c/n)^2$ of the transactions. 
\autoref{tab:ases} shows the Tor relay statistics observed for the top five autonomous systems. 
\\
If the path from the sender to the anonymity network and the path from the anonymity network to destination traverse separate ASes, then observing the transactions to reveal private data can be prevented when the ASes do not collude. Furthermore, Tor has multiple relays participating in a routing circuit belonging to the same AS. For instance, Tor has three nodes in AS 23504 (Speakeasy DSL), which harms the anonymity of traffic. To prevent deanonymization attacks in Tor traffic, it must ensure that the ASes of the nodes present in a circuit path are not colluding. This property could be achieved by limiting the number of Tor relays present in an AS, which can be hard to achieve since it depends on the volunteers who run the relays.
\\
A proposed solution to achieve location independence can involve considering the actual AS of a node, not simply its IP address while selecting a circuit. Previous studies suggest that nodes in edge networks (e.g., cable modem and DSL providers, universities, etc.) are likely to traverse the same AS on both the inbound and outbound paths to those nodes. Far-flung node locations that provide geographical diversity, such as nodes in Asia, are likely to reduce location independence because such nodes do not typically have diverse AS-level connectivity. Ensuring location diversity for providing anonymity does not simply mean that the routers must be located at a different geographic location. It means that the routers must be present in ASes that are connected to a large number of other ASes (they do not lie on the edge of the network). The best place for new nodes is likely to be in ASes that have a high degree—that is, those that connect to a large number of other ASes. Ironically, the ASes with the highest degree tend to be tier-1 ISPs themselves; thus placing one node in each tier-1 ISP and building mixed paths between those nodes may be the best strategy for increasing location independence.

\begin{table*}[hbt!]\centering
\begin{tabular}{l|l|l|l|l|l|l|l|l}

 \textbf{\begin{tabular}[c]{@{}l@{}}Autonomous\\ System\end{tabular}} &
  \textbf{\begin{tabular}[c]{@{}l@{}}Consensus\\ Weight\end{tabular}} &
  \textbf{\begin{tabular}[c]{@{}l@{}}Advertised\\ Bandwidth\end{tabular}} &
  \textbf{\begin{tabular}[c]{@{}l@{}}Guard\\ Probability\end{tabular}} &
  \textbf{\begin{tabular}[c]{@{}l@{}}Middle\\ Probability\end{tabular}} &
  \textbf{\begin{tabular}[c]{@{}l@{}}Exit\\ Probability\end{tabular}} &
  \textbf{Relays} &
 \textbf{Guard} &
  \textbf{Exit} \\ \hline
\textbf{\begin{tabular}[c]{@{}l@{}}Hetzner Online\\ GmbH 
(AS24940)\end{tabular}} &
  12.6683\% &
7363.92 MiB/s &
19.2642\% &
17.9991\% &
0.0625\% &
369	&
247	&
3  \\
\textbf{\begin{tabular}[c]{@{}l@{}} 
OVH SAS\\(AS16276)
\end{tabular} } &
  	8.7602\% &
  	5830.72 MiB/s &
  	12.1351\% &
  	11.0134\% &
  	2.8116\% &
  	399	 &
  	283	&
  	45 \\ 
\textbf{\begin{tabular}[c]{@{}l@{}} 
CIA TRIAD\\SECURITY LLC
\end{tabular} } &
   6.7848\% &
   4125.32 MiB/s &
   0.0000\% &
   0.0000\% &
   21.1266\% &
   191 &
   69 &
   191 \\ 
\textbf{\begin{tabular}[c]{@{}l@{}} 
ONLINE S.A.S.\\(AS12876)
\end{tabular} } &

      5.0770\% &
      3233.15 MiB/s &
      8.0394\% &
      5.8607\% &
      1.1174\% &
      196 &
      167 &
      16 \\
\textbf{\begin{tabular}[c]{@{}l@{}} 
PONYNET\\(AS53667)
\end{tabular} } &
        
    	4.8461\% &
    	3723.71 MiB/s &
    	1.5086\% &
    	1.2846\% &
    	12.1378\% &
    	405 &
    	358 &
    	321 \\
\end{tabular}
\caption{Tor Statistics Observed in Top 5 ASes}
\label{tab:ases}
\end{table*}


\section{Conclusion}
From its origins as a research project and early use mostly by the technical community, Tor has evolved into being a usable tool with significant societal impact. Recent events in the past month alone highlight the continued relevance of Tor to such studies. It was revealed this past month that a threat actor was running thousands of Tor relays in an attempt to deanonymize parts of the network. Also this past month, Russia attempted to prevent the use of the network by blocking access to the \texttt{torproject} website and directing ISPs to block proxy access\cite{cimpanu21}\cite{ikeda21}. Measuring Tor, therefore, poses interesting questions and challenges outside of the solely technical. 

In this paper we sought to understand and explain trends in some different aspects of Tor over the near-twenty years it has been in use. We examined previous measurement studies of Tor, the unique considerations necessary when studying such a tool, and the impact of Tor on real-world issues of censorship, and security issues of location diversity.


\bibliographystyle{unsrt}
\bibliography{refs}
\end{document}